\documentclass[prl,twocolumn]{revtex4}
\usepackage{times}
\usepackage[pdftex]{graphicx}
\usepackage[pdftex]{epsfig}
\usepackage{amsmath}
\usepackage{pslatex}
\usepackage{amssymb}
\usepackage{amsfonts}
\usepackage{color}
\usepackage{wrapfig}
\usepackage{eucal}
\usepackage{hhline}
\usepackage{threeparttable}
\usepackage{supertabular}
\usepackage{multirow}
\usepackage{tabularx}
\usepackage{hyperref}

\newcolumntype{Y}{>{\centering\arraybackslash}X}

\begin{document}

\title{Characterization of radiation pressure and thermal effects in a nanoscale optomechanical cavity}

\author{Ryan M. Camacho, Jasper Chan, Matt Eichenfield, and  Oskar Painter}
\affiliation{Thomas J. Watson Sr. Laboratories of Applied Physics, California Institute of Technology, Pasadena, California 91125}

\begin{abstract}
Optical forces in guided-wave nanostructures have recently been proposed as an effective means of mechanically actuating and tuning optical components.  In this work, we study the properties of a photonic crystal optomechanical cavity consisting of a pair of patterned $\rm{Si_3N_4}$ nanobeams.  Internal stresses in the stoichiometric $\rm{Si_3N_4}$ thin-film are used to produce inter-beam slot-gaps ranging from $560$ to $40$~nm.  A general pump-probe measurement scheme is described which determines, self-consistently, the contributions of thermo-mechanical, thermo-optic, and radiation pressure effects.  For devices with $40$~nm slot-gap, the optical gradient force is measured to be $134$~fN per cavity photon for the strongly coupled symmetric cavity supermode, producing a static cavity tuning greater than five times that of either the parasitic thermo-mechanical or thermo-optic effects.             
\end{abstract}

\maketitle

Radiation pressure forces have recently been studied in the context of mechanically compliant optical microcavities for the sensing, actuation, and damping of micromechanical motion\cite{Kippenberg082,ref:Favero2}.  A wide variety of cavity geometries have been explored, from Fabry-Perot cavities with movable internal elements or end-mirrors\cite{ref:Dorsel,Cohadon99,Gigan06,Kleckner06,Corbitt07,ThompsonJD08}, to monolithic whispering-gallery glass microtoroids\cite{Schliesser06}.  Nanoscale guided-wave devices have also been studied due to their strong optomechanical coupling resulting from the local intensity gradients in the guided field\cite{ref:Povinelli051,Notomi06,Eichenfield07,Li08,Eichenfield09,Anetsberger09,LinQ09}.  In addition to radiation pressure forces, there exists in each of these cavity geometries competing thermally driven effects, a result of optical absorption.  Thermally induced processes include strain-optical, thermo-optical, and a variety of thermo-mechanical effects (early measurements of radiation pressure, for instance, were plagued by thermo-mechanical ``gas action'' effects\cite{Nichols1901}).  In many cases (but not all\cite{ref:Hohberger1,ref:Ilic1}), thermal effects can be neglected at the high frequencies associated with micromechanical resonances\cite{Schliesser06}; however, for static cavity tuning thermal effects may play a significant, if not dominant, role.  Calibration of thermal effects is important not only in identifying the contribution of pure radiation pressure effects, but also in understanding the parasitic local heating processes, which in the realm of quantum optomechanics may limit optical cooling methods\cite{Marquardt07,Wilson-Rae07}, or for tunable photonics applications\cite{Rakich07,Li08,RosenbergJ09} where deleterious inter-device thermal coupling may arise.

In this work we describe the characterization of the low frequency (static) optical and thermal effects in a nanoscale photonic crystal cavity.  This so-called zipper cavity\cite{ChanJ09,Eichenfield09,Deotare09} consists of a matched pair of $\rm{Si_3N_4}$ nanobeams, placed in the near-field of each other, and patterned with a one-dimensional (1D) array of air holes.  The resonant optical modes of the zipper cavity\cite{ChanJ09} consist of manifolds of even (bonded) and odd (anti-bonded) symmetry supermodes of the dual nanobeams, localized along the long-axis of the beams to a central \emph{defect} in the photonic lattice.  The strength of the gradient optical force applied to the beams is exponentially dependent upon the inter-beam slot gap ($s$), and parameterized by a dispersive coupling coefficient, $g_{\text{OM}} \equiv 2 \left(\partial\omega_{c}/\partial s\right)$, where $\omega_{c}$ is the gap-dependent optical cavity resonance frequency.  Utilizing the connection between cavity mode dispersion and applied optical force, a pump-probe scheme is described below to accurately quantify the optomechanical, thermo-optic, and thermo-mechanical contributions to the static tuning of the bonded and anti-bonded modes versus internal cavity photon number.              
 
\begin{figure*}[t]
\includegraphics[width=1.6\columnwidth]{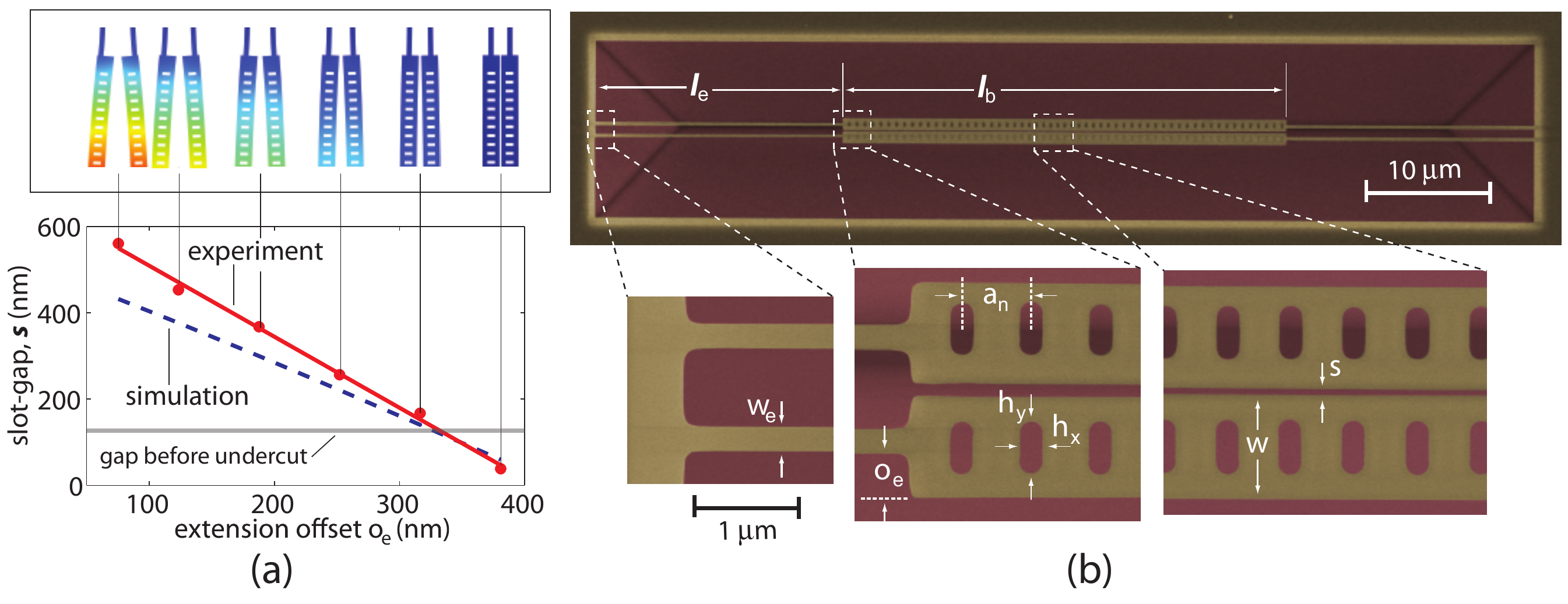}
\caption{(a) Slot width vs. extension offset, simulation and experiment.  Support extensions placed near the outside of the beam width cause outward bowing, but inward bowing when placed near the inside. (b) Scanning electron microscope images of a device with support extensions placed just inside of center, causing slight inward bowing resulting in approximately a slot-gap of $s=40$~nm at the cavity center.  The device parameters that are common amongst all the devices tested in this work are: nominal lattice constant $a_{n}=590$~nm, extension width $w_{e}=211$~nm, hole width $h_{x}=190$~nm, hole height $h_{y}=416$~nm, beam width $w=833$~nm, beam length $l_{b}=18.4$~$\mu$m, extension length $l_{e}=33.1$~$\mu$m.  In order to vary the slot-gap size from $40$-$560$~nm, the extension offset was varied between $o_{e}=75$-$380$~nm.}
\label{fig1_devices}
\end{figure*}

The zipper cavity devices studied here were fabricated from optically thin ($t = 400$ nm) stoichiometric silicon nitride (Si$_3$N$_4$), deposited using low-pressure-chemical-vapor-deposition on a silicon wafer.  The deposition process results in residual in-plane stress in the nitride film of approximately $\sigma \sim 1$ GPa\cite{ref:Verbridge1}.   Electron-beam lithography is used to pattern zipper cavities with beams of length $l_b = 33.1$~$\mu$m, width of $w =833$~nm, and an inter-beam spacing of $s = 120$ nm.  The beams are clamped at each end using extensions with length equal to $l_e = 18.4$~$\mu$m and width of $w_e=211$~nm a shown in Fig.~\ref{fig1_devices}(b).  The nanobeam and photonic crystal hole pattern are transferred into the Si$_3$N$_4$ film using a $\rm{C_4F_8/SF_6}$ plasma etch. The underlying Si substrate is selectively etched using KOH, releasing the patterned beams.  The devices are dried using a critical point $\rm{CO_2}$ drying process to avoid surface-tension-induced adhesion of the nanobeams.

The internal stress of the Si$_3$N$_4$ thin-film is used to create a range of slot-gaps.  This is achieved through misalignment of the support extensions and central nanobeams (Fig.~\ref{fig1_devices}(b)), breaking the symmetry of the internal stress along the length of the beams.  As shown by the finite-element-method (FEM) simulations and device measurements plotted in Fig.~\ref{fig1_devices}(a), extensions placed near the outside edge of the beams cause outward bowing of the beams, while extensions placed near the inside edge cause inward bowing.  By varying the lateral offset of the extension from $o_e= 75$ to $380$~nm, zipper cavities with slot-gaps at the cavity center ranging from $40$-$560$ nm are created.  Slot-gaps smaller than $40$~nm could not be stably produced without the nanobeams sticking together at points of nanometer-scale roughness in the inner sidewall of the beams. 

\begin{figure}[t]
\includegraphics[width=\columnwidth]{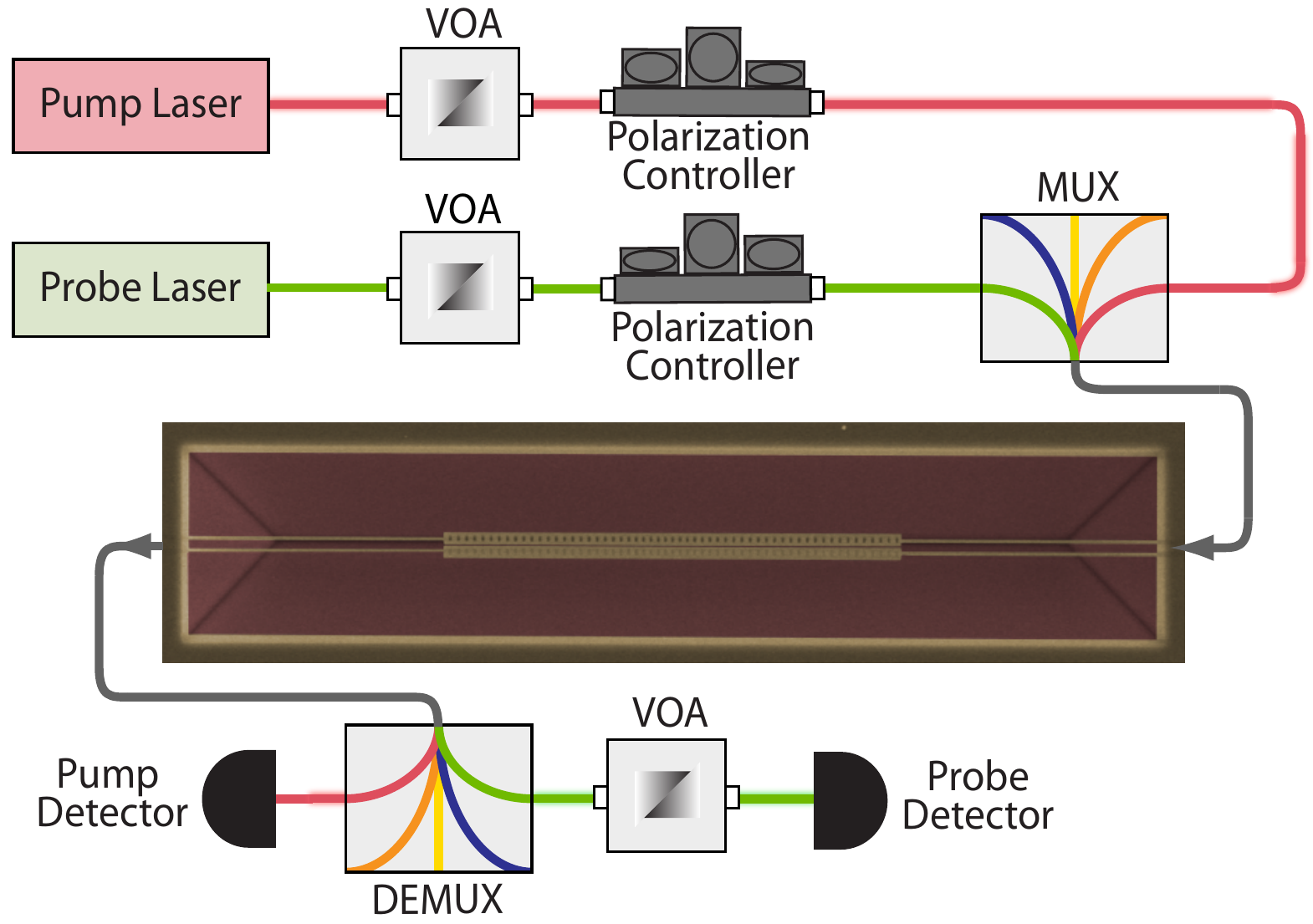}
\caption{Experimental setup for optical testing.  Two separate lasers (pump and probe) with independent power control, via variable optical attenuators (VOA), and polarization control are combined into a fiber taper waveguide placed in the near-field of the photonic crystal cavity.  Cavity transmission at both the pump and probe wavelengths are multiplexed and demultiplexed using a matched set of fiber-based filters and separately monitored using calibrated photodetectors. }
\label{fig2_expsetup}
\end{figure}

Optical spectroscopy of the zipper cavity modes is performed using the experimental setup shown in Fig. \ref{fig2_expsetup}.  In this setup, laser light from a bank of tunable external-cavity diode lasers covering the $1400$-$1625$~nm wavelength band is coupled into an optical fiber taper nanoprobe.  Using precision motorized stages, the tapered fiber can be controllably placed into the near-field of a zipper cavity\cite{ref:MichaelCP1}, allowing for evanescent excitation and detection of resonant modes.  Polarization of the laser field is adjusted using a fiber polarization controller, and optimized for coupling to the high-$Q$ TE-like modes of the zipper cavity\cite{ChanJ09} (i.e., dominant electric field polarization in the plane of the device).

\begin{figure*}[t]
\includegraphics[width=1.3\columnwidth]{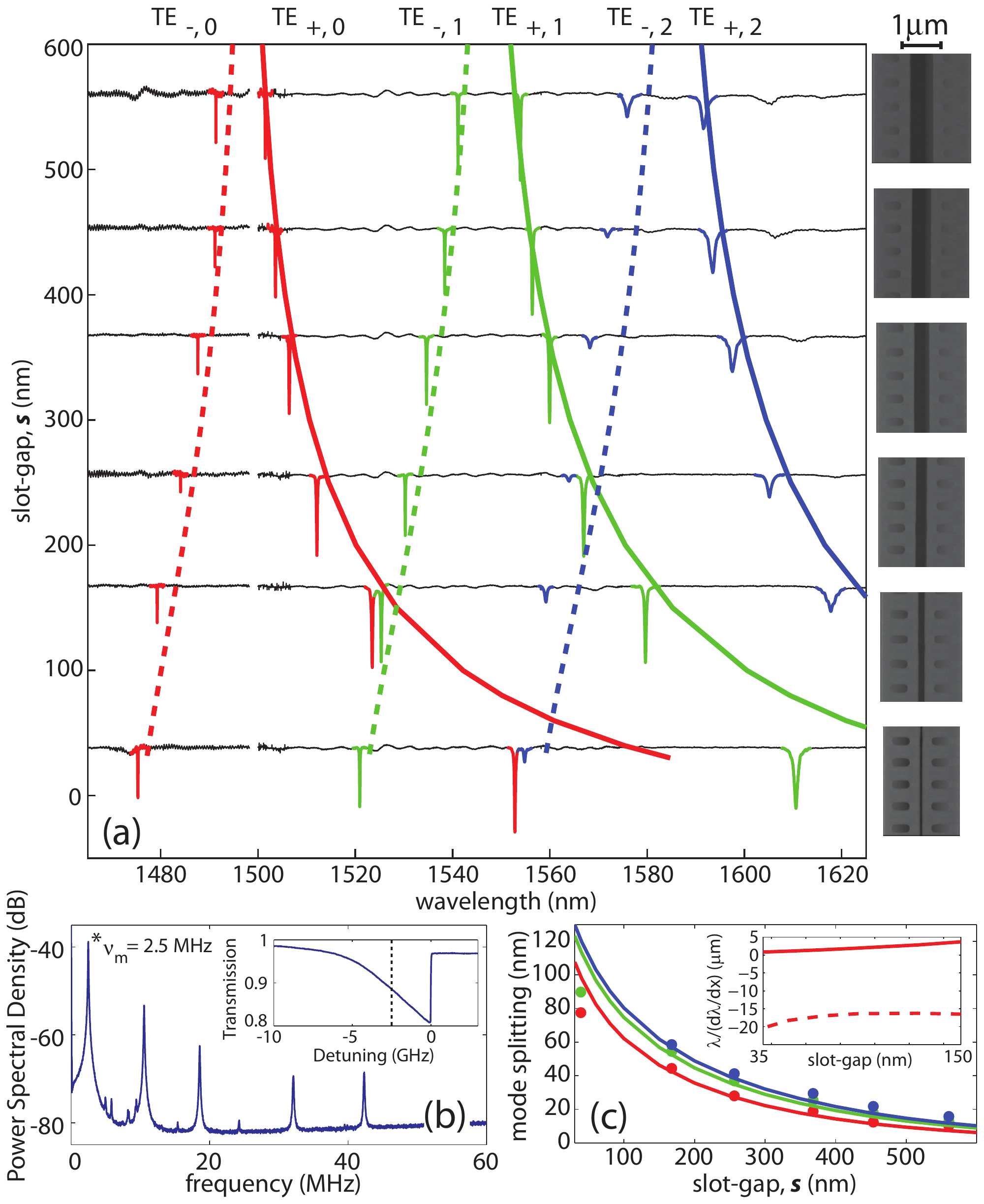}
\caption{(a) Experimentally measured optical transmission as a function of wavelength in an array of six devices (SEM images of the central cavity region on right), each with a different slot-gap.  Overlayed are FEM simulations of the cavity mode dispersion versus gap size, where solid curves are for the bonded modes, dashed curves for the anti-bonded modes, and the color of the curve matches the highlighting applied to the different mode orders (red=TE$_{\pm,0}$, green=TE$_{\pm,1}$, blue=TE$_{\pm,2}$).  (b) Measured RF spectrum of the TE$_{+,0}$ mode for the largest gap ($s=560$~nm) zipper cavity.  Inset shows the optical transmission as a function of detuning as the pump laser is swept across the cavity resonance (dashed vertical line indicates detuning for RF spectrum measurement). (c) Plot of the FEM-simulated and experimentally measured bonded and anti-bonded mode splitting versus slot-gap.  The inset shows the derivative of the simulated dispersion curve in the small slot-gap region (solid curve for bonded mode, dashed curve for anit-bonded mode).}
\label{fig3_resonances}
\end{figure*}

Figure \ref{fig3_resonances}(a) shows a series of wavelength scans from an array of nominally identical zipper cavities, each with slightly differing slot-gap due to variation in the lateral extension alignment.  Scanning electron microscope images of the central region of each zipper cavity, from which the slot-gap is measured, are shown to the right of each wavelength scan.  Even (bonded) and odd (anti-bonded) parity supermodes are identified by stepping the taper across the width of the zipper cavity and noting the lateral spatial symmetry of the mode coupling\cite{Eichenfield09}.  Three distinct pairs of modes are identified in the wavelengh scans, with the fundamental longtitudinal cavity modes (TE$_{\pm,0}$) occurring at shorter wavelengths and the higher-order modes shifted to longer wavelengths\cite{ChanJ09}.  Radio frequency (RF) spectra of the optical transmission intensity (see Fig. \ref{fig3_resonances}(b)) were performed to verify the presence of micromechanical oscillation and free movement of the nanobeams even for the smallest of slot-gaps. 

Overlayed on the wavelength scan plots of \ref{fig3_resonances}(a) are theoretical dispersion curves generated using FEM simulations of the optical properties of the zipper cavity (a single in-plane scaling factor of $5 \%$ was used to match the TE$_{+,0}$ resonant wavelength for the largest ($560$~nm) slot-gap, which is within the accuracy of our SEM calibration).  Figure \ref{fig3_resonances}(c) shows a plot of the measured and simulated splitting of the fundamental bonded and anti-bonded modes versus the SEM-measured slot gap.  Good correspondence is found for all but the smallest ($40$~nm) slot gap, in which the theoretical curve shows significantly more dispersion for the bonded modes.  The source of this discrepancy is not fully understood, but may be due to other effects such as the dispersive nature of the refractive index of the Si$_3$N$_4$ material itself.  As discussed below, the local sensitivity of each mode to beam displacement is consistent with the measured $40$~nm slot gap.               

In order to understand the various actuation methods of the zipper cavity, it is useful to consider the dispersive nature of the two supermode mode types, and their relation to the magnitude and direction of the optical force.  In an adiabtic limit\cite{Law1994,ref:Povinelli051}, the radiation pressure force can be related to the gradient of the internal optical cavity energy, $F_{\text{OM}} = -\partial (N\hbar\omega_{c})/\partial \alpha = -N\hbar g_{\text{OM}}$, where $N$ is the stored photon number and $\alpha$ is a displacement factor related to the movement of the nanobeams (here we choose $\alpha$ to be equal to one half the slot-gap size).  As can be seen in the measured and simulated dispersion curves of Figure \ref{fig3_resonances}(c), the symmetric bonded modes with large field strength in between the beams, tunes to the red with shrinking slot-gap (i.e., $g_{\text{OM}}$ for the bonded mode is positive for $\alpha=s/2$).  The direction of the optical force for photons stored in a bonded mode thus tends inwards, pulling the beams together (this is a result of the fact that in order to perform mechanical work on the beams, the stored photons must lose energy through a reduction in their frequency).   In the case of the odd parity anti-bonded cavity modes, the resonance frequency decreases with increasing slot-gap, resulting in a negative $g_{\text{OM}}$ and an optical force that pushes the beams apart.      

Owing to the different dispersive character of the bonded and anti-bonded cavity modes, and the different directions in which cavity photons of each mode type apply forces to the two beams, pure radiation pressure effects actuate and tune the cavity modes in a unique and distinctive manner.  This should be contrasted with thermo-optic and thermo-mechanical effects.  The thermo-optic tuning of the cavity modes is related to the change in refractive index of the cavity material with temperature ($\partial n/\partial T \approx 1.9 \times 10^{-5}$ K$^{-1}$ for Si$_3$N$_4$\cite{Eichenfield09}).  The manifold of bonded and anti-bonded cavity modes thus tend to uniformly red shift with increasing temperature due to the thermo-optic effect.  Increased temperature in the local cavity region of the beams, due to optical absorption, not only increases the refractive index of the clamped beams, but may also produce non-uniform thermal expansion and significant strain in the structure.   Unlike the radiation pressure force, the resulting thermo-mechanical force is only dependent upon the temperature rise, and thus actuates the zipper cavity beams in the same manner independent of which cavity mode is being driven.  Owing to the different dispersive nature of the cavity modes, however, the sign of the thermo-mechanical tuning will depend upon the type of supermode being considered.    

\begin{figure}[t]
\includegraphics[width=\columnwidth]{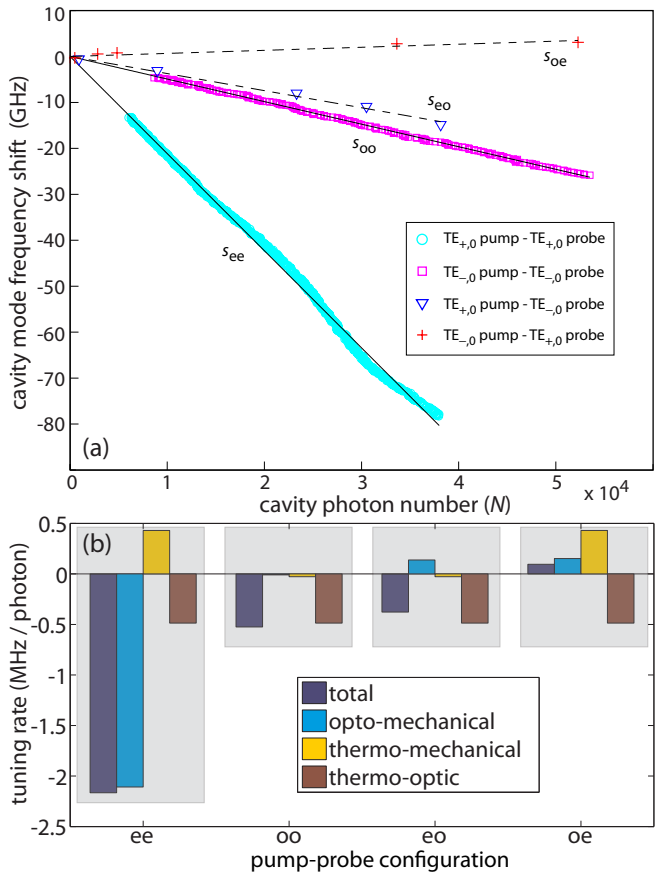}
\caption{(a) Measured tuning curves for the four pump-probe configurations.  (b) Opto-mechanical, thermo-mechanical, and thermo-optic tuning contributions as determined from a fit of the model coefficients.}
\label{fig4_tuningplots}
\end{figure}

We perform measurements of the different cavity mode tuning mechanisms using a two laser pump and probe scheme.  In this scheme, a strong pump beam is coupled into either the fundamental bonded (TE$_{+,0}$) or anti-bonded (TE$_{-,0}$) cavity mode.  The pump laser frequency is swept across the cavity mode resonance producing a bistability curve such as that shown in Fig. \ref{fig3_resonances}(b).  A fit to the bistability curve yields the self-mode tuning versus dropped power into the cavity.  The dropped cavity power is then converted into an internal photon number via the intrinsic $Q$-factor of the cavity mode\cite{ref:Barclay7} (measured to be $Q_{i}\approx 6\times 10^4$ for both the fundamental bonded and anti-bonded modes using a calibrated fiber Mach-Zender interferometer).  As the pump beam is scanned across one of the modes, a weak probe beam is scanned across the other.  A fit to the Lorentzian lineshape of the probe scan yields the cross-mode tuning versus dropped pump power.  The measured self and cross tuning curves of the smallest slot-gap ($40$~nm) zipper cavity, with the strongest optomechanical coupling, are shown in Fig. \ref{fig4_tuningplots}(a).  

These measurements yield four tuning slopes $s_{ij} = \Delta\omega_j/N_i$, where subscripts $i,j \in \{e,o\}$ label the cavity mode ($e$ for the even bonded mode, $o$ for the odd anti-bonded mode), $N_{i}$ is the stored cavity photon number of the pump mode, and $\Delta\omega_j$ is the induced frequency shift in the $j$th mode.  The tuning slopes can be related to the four coefficients describing the opto-mechanical, thermo-mechanical, and thermo-optic per-photon forces/dispersion:

\begin{equation}
s_{ij} = \left(\frac{-\hbar(g_i +g_{tm})}{k}\right)g_j + c_{to}.
\label{Eq:tuning_slopes}
\end{equation}

\noindent Here $g_i$ and $g_j$ are the optomechanical coupling coefficients of the pump and probe cavity modes, respectively, $g_{tm}$ is an effective thermo-mechanical force coefficient, $c_{to}$ is a thermo-optic tuning coefficient, and $k$ is the spring constant associated with differential in-plane motion of the zipper cavity nanobeams.  

Solving the set of coupled equations for the four cavity tuning coefficients yields (for $\alpha \equiv s/2$, i.e., $g_{e} > 0$, $g_{o} < 0$),

\begin{subequations}
\begin{eqnarray}
g_e &=& -\xi\sqrt{\frac{k}{\hbar}}\Delta_e, \\
g_o &=& \xi\sqrt{\frac{k}{\hbar}}\Delta_o, \\
g_{tm} &=& \xi\sqrt{\frac{k}{\hbar}}\Delta_c, \\
c_{to} &=& -\xi^2(s_{ee}s_{oo} - s_{eo}s_{oe}),
\end{eqnarray}
\label{Eq:tuning_coefficients}
\end{subequations}

\noindent where $\xi \equiv 1/\sqrt{-(\Delta_{e}+\Delta_{o})}$, and we have defined relative tuning slopes $\Delta_e = s_{ee} - s_{oe}$,  $\Delta_o =s_{oo} - s_{eo}$, and $\Delta_c = s_{eo} - s_{oe}$.  

\noindent From the measured and fit tuning slopes shown in Fig. \ref{fig4_tuningplots}(a), we extract the following values for the optomechanical and thermal coefficients of the smallest slot-gap zipper cavity: $\sqrt{\hbar/k}g_e = 1.44 \pm 0.04$, $\sqrt{\hbar/k}g_o=-0.083 \pm 0.021$, $\sqrt{\hbar/k}g_{tm} = -0.29 \pm 0.01$, and $\sqrt{-c_{to}} = 0.68 \pm 0.02$, all in units of $\rm{\sqrt{MHz/photon}}$.  The uncertainty in the measured coefficients is dominated by systematic error in the internal cavity photon number ($\sim 5\%$ due to uncertainty in cavity $Q$ and optical input power).  The resulting contributions from this model to the cavity mode tuning are shown graphically in Fig. \ref{fig4_tuningplots}(b).  

For an effective mechanical spring constant of $k=13.18$~N/m (estimated from the motional mass $m_{x}\approx53$~pg\cite{ChanJ09} and the fundamental in-plane mechanical mode frequency of $\Omega_{M}/2\pi = 2.5$~MHz), the corresponding optomechanical coupling coefficients are $g_{e}/2\pi = 202.4 \pm 5.6$~GHz/nm and $g_{o}/2\pi = -11.63 \pm 2.9$~GHz/nm.  These should be compared with the theoretical values estimated from FEM simulation at the SEM-measured slot-gap for this device ($s=40$~nm), which are $g_{e}/2\pi = 208$~GHz/nm and $g_{o}/2\pi = -11.4$~GHz/nm.  A similar zipper cavity structure was also tested that did not have the thin beam extensions and with ten times the mechanical spring constant.  The resulting cavity tuning rate in this structure is dominated by the thermo-optic effect, and is measured to be $-0.69$~MHz/photon, similar to the pump-probe measured value of $c_{to}=-0.47 \pm 0.03$~MHz/photon for the device studied in Fig.~\ref{fig4_tuningplots}.  From measurements of the cavity tuning rate versus substrate temperature of the stiff structure ($-1.87$~GHz/K), the estimated temperature rise rate in the cavity of Fig.~\ref{fig4_tuningplots} (using the pump-probe measured $c_{to}$) is $\beta=2.5\times 10^{-4}$~K/photon.  For the largest pump powers used here this corresponds to a maximum cavity temperature rise of about $14$~K.  FEM simulations of the thermo-mechanical properties of the cavity structure show an increase of the inter-beam slot-gap per Kelvin of temperature rise in the central cavity region equal to $\delta s_{tm}=0.021$~nm/K.  This yields a simulated change in beam displacement factor per cavity photon of $\delta\alpha_{tm}= (\delta s_{tm}/2)\beta \approx 2.6$~fm/photon, or a thermo-mechanical force coefficient of $\tilde{g}_{tm}=-(k/\hbar)\delta\alpha_{tm} \approx -50$~GHz/photon, in good correspondence with the pump-probe measured value of $g_{tm} = -40.8 \pm 1.4$~GHz/photon.
            
In conclusion, we have introduced a pump-probe method to completely characterize opto-mechanical, thermo-mechanical, and thermo-optic contributions to the static tuning of a zipper optomechanical cavity.  More generally, this approach is useful for guided-wave devices in which near-field coupling between two (or more) elements is used to produce the optical gradient force, and for which the different supermodes of the composite structure introduce similar thermal and differing radiation pressure effects.  We demonstrate the method experimentally using a strongly opto-mechanically coupled device, fabricated using a technique that allows for fine control over a nanoscale slot-gap which sets the magnitude of the optical gradient force.  Using a device with a measured $40$~nm slot-gap, we characterize the various thermal and optical tuning mechanisms and find that the static tuning of the device is dominated by the optical gradient force, with optomechanical coupling coefficient $g_{\text{OM}}\approx 200$ GHz/nm (equivalent to $0.13$pN/photon).  Beyond cavity optomechanics, the creation of ultra-small (tens of nanometer) slot gaps in photonic devices is interesting for cavity qauntum electrodynamics\cite{ref:Robinson1,ChanJ09}, nonlinear optics\cite{Koos09}, and sensing\cite{Robinson08}, due to the extremeley large per-photon electric field strengths that build up in such nano-slots.   


\end{document}